\documentclass[showpacs,twocolumn,aps,amssymb,preprintnumbers,superscriptaddress,prc,floatfix]{revtex4}

\usepackage{epsfig,dcolumn}

\usepackage{graphicx,bm}

\usepackage{bm,color}

\usepackage{hyperref}

\newcommand{\be}{\begin{equation}}

\newcommand{\ee}{\end{equation}}

\newcommand{\ba}{\begin{eqnarray*}}

\newcommand{\ea}{\end{eqnarray*}}

\begin{document}

\title{The land of deformation south of $^{68}$Ni}

\author{S.~M.~Lenzi}

\affiliation{Dipartimento di Fisica dell'Universit\`a and INFN, Sezione di Padova, I-35131 Padova, Italy}

\author{F. Nowacki}

\affiliation{IPHC, IN2P3-CNRS et Universit\'e de Strasbourg, 
F-67037 Strasbourg, France}

\author{A. Poves}

\affiliation{Departamento de F\'isica Te\'orica e IFT-UAM/CSIC, 
Universidad Aut\'onoma de Madrid,  E-28049 Madrid, Spain}

\author{K. Sieja}

\affiliation{IPHC, IN2P3-CNRS et Universit\'e de Strasbourg, 
F-67037 Strasbourg, France}

\begin{abstract}

We study the development of  collectivity 
in the neutron-rich nuclei around $N=40$, where  experimental and theoretical
evidences suggest a rapid shape change from the spherical to the rotational regime,
in analogy to what happens at the {\it island of inversion} surrounding $^{31}$Na.
Theoretical calculations are performed within the interacting shell model framework
in a large valence space, based on a $^{48}$Ca core which encompasses 
the full $pf$ shell for the protons and the $0f_{5/2}$, $1p_{3/2}$, $1p_{1/2}$, $0g_{9/2}$ and  $1d_{5/2}$ orbits
for the neutrons. The effective interaction is based on a G-matrix obtained from a 
realistic nucleon-nucleon potential whose monopole part is
corrected empirically to produce effective single particle energies compatible with the
experimental data.  
We find  a good agreement between the theoretical results and the available
experimental data. We predict the onset of deformation at different neutron numbers
for the various  isotopic chains.  The maximum collectivity occurs  in the chromium isotopes,
where the large deformation regime starts already at $N=38$.
The shell evolution responsible for the observed shape changes is discussed
in detail, in parallel to the situation in  the $N=20$ region.

\end{abstract}

\pacs{21.60.Cs, 21.10.--k, 21.10.Re}
\keywords{ Shell model, Effective interactions,
 Spectroscopy, Level schemes and transition probabilities.}

\date{\today}

\maketitle

\section{Introduction}

In the last decades, more and more  experimental evidence 
has been accumulated establishing the breaking of the shell closures 
known at the stability valley when approaching the drip lines,
mainly at the very neutron-rich side. The first example of an unexpected 
disappearance of a shell closure ($N=8$) was found in $^{11}$Be, whose 
ground state is an intruder $1/2^+$ located 320~keV below the "natural"
 $0\hbar\omega$ $1/2^-$ state \cite{Alburger1964}. However, the true relevance of this finding
was not recognized till  many years later, and even now it is shadowed by the fame of its more
neutron-rich isobar$^{11}$Li.  
In the $sd$ shell, the expected $N=20$ semi-magic  neutron rich 
nuclei turned out to be actually well deformed \cite{Thibault1975,Detraz1983}.   
The presence of deformed ground states in the nuclei around $^{32}$Mg and the
breaking of the $N=20$ shell closure in this region
has been extensively studied experimentally and theoretically (see \cite{Caurier:2004gf,brown:rev,Otsuka2001319} 
for shell model reviews). The aim of these studies is to map the limits of the so-called
 {\it island of inversion}, i.e. the region of nuclei
where the strong quadrupole correlations overcome the spherical mean-field gaps,
favoring  energetically the deformed intruders, which often become  ground states.
Indeed, another basic aim  is to understand microscopically the dynamics responsible
for these shape transitions. 

The question of the persistence of the $N=40$ harmonic oscillator closure in neutron-rich nuclei
comes in naturally in this context. The high lying $2^+$ state
observed in $^{68}$Ni and its low $B(E2; 2^+\rightarrow0^+)$ value are 
the result of the relatively large energy gap separating 
the $pf$ and 0$g_{9/2}$ orbitals \cite{Sorlin2002}. However,  this gap gets reduced (or even disappears)
when protons are removed from $^{68}$Ni: The nucleus $^{66}$Fe, with only
two protons less, shows a sudden change in nuclear structure
with an increased collectivity manifested via its very low lying $2^+$ state.
Along the iron chain, indications for a collective behavior come
from the systematics of the $2^+$ states~\cite{Hannawald} as well as from 
the recent measurement of the $B(E2)$ values in $^{64,66}$Fe~\cite{Ljungvall, Rother}.
The evolution of the $B(E2)$
values in iron isotopes points to a sudden increase of collectivity when approaching $N=40$. 
Only very recently the first measurement of the excited levels 
in $^{64}$Cr has been reported~\cite{Gade10}.  This is the nucleus where  
theoretical calculations predicted the lowest lying
$2^+$ level in the region~\cite{Caurier2002, Sor03, Kaneko08}.  
The measured $2^+$ state energy  agrees within 100 keV with those theoretical predictions.
The sudden shape change at $N=40$ challenges the theoretical models,
which have to account for a particularly rapid shell evolution
responsible for these effects. 
Recent beyond 
mean-field HFB+GCM calculations with the Gogny force, reported in Ref.~\cite{Gaudefroy:2009zz},
show an increase of collectivity towards the proton drip line. The spherical neutron 
single-particle energies obtained in the HFB approach reveal almost no variation of
the $N=40$ gap with the proton number between $Z=20$ and $Z=28$. As a consequence,
only a moderate collectivity in the iron  and chromium chains is found, without visible structure changes
between them. The collectivity at $N=40$ has been previously 
a subject of many shell model studies using different valence spaces and interactions
 \cite{Caurier2002, Sor03, Lun07, Kaneko08, Gade10}.
In particular, it has been shown that the shell model calculations  
using the $0f_{5/2}$, $1p_{3/2}$, $1p_{1/2}$, $0g_{9/2}$ neutron orbits (the $fpg$ valence space)
and realistic interactions~\cite{Sor03}, can reproduce rather well the level schemes 
of $^{62,64}$Fe, but fail to do so for the $2^+$ state of  $^{66}$Fe~\cite{Lun07}.  
To reproduce the large quadrupole collectivity in this mass region, the inclusion of the neutron $1d_{5/2}$ 
orbital is needed, as first surmised in Ref.~\cite{Caurier2002} and confirmed
recently in Ref.~\cite{Ljungvall}. This can be explained in terms of the quasi-SU3 approximate symmetry: 
In this framework, the deformation  is  generated by the interplay between the quadrupole 
force and the central field in the subspace consisting on the lowest $\Delta j = 2$ 
orbitals of a major shell~\cite{Zuker1995}. 

In this work we discuss in detail how the sudden onset of collectivity 
can be interpreted in terms of shell model calculations in large model spaces
and we look for the similarities in the deformation-driving mechanisms at $N=20$ and $N=40$
when approaching the neutron drip line. 
We present novel calculations for the $N=40$ region, in a model space
including the $pf$-shell for protons and the $0f_{5/2}$, $1p_{3/2}$, $1p_{1/2}$, $0g_{9/2}$ and $1d_{5/2}$ orbits for neutrons. 
We discuss first our model space and
interaction as well as the computational details in Sec. \ref{vs}.
The shape evolution along the $N=40$ line is studied in Section \ref{n40}.
Next, we illustrate the development of deformation along the isotopic
chains of iron and chromium in Sec. \ref{ich} and we discuss the structure
of the nickel isotopes. Conclusions are given in Sec. \ref{con}.

\section{The valence space\label{vs}}

The physics of the $N=40$ nuclei has been discussed in the recent past  both
experimentally and theoretically 
\cite{brown:rev,Hannawald,Ljungvall,Gade10,Caurier2002,Sor03,Kaneko08,Gaudefroy:2009zz,Lun07}. In $^{68}$Ni,
the $N=40$ harmonic oscillator shell closure
is somewhat weakened, however, whereas some of its properties 
seem consistent with a superfluid behavior, others may point to a double 
magic character. Below  $^{68}$Ni, the iron and chromium 
isotopic chains are characterized by
an open proton shell  which favors the development of  quadrupole correlations.  
Such a situation occurs also for the $N=20$ nuclei in the so called {\it island of inversion}
around $^{32}$Mg \cite{SDPF-M}: the protons  occupy the $N=2$ harmonic oscillator 
shell and the neutrons may lie either in  the  $N=2$ harmonic oscillator shell (normal filling)
or in the $1p_{3/2}$ and $0f_{7/2}$ orbitals from the $N=3$ major 
shell (intruder configurations). The latter orbits  form a  quasi-SU3 block \cite{quasi-SU3}
which enhances the quadrupole correlations.
Extended shell model calculations by the Tokyo group,  with the SDPF-M
effective interaction \cite{SDPF-M},  predict the dominance of different type 
of configurations (0p0h, 1p1h, 2p2h) for the ground states of the nuclei with
$18 \le N \le 22$ and  $10 \le Z \le 14$, delineating the
contours of the island of inversion.

To assess if such a scenario could develop around  $^{68}$Ni,
we adopt here a model space based on a $^{48}$Ca core which
comprises  the $pf$-shell  for protons and the 1p$_{3/2}$, 
$1p_{1/2}$, $0f_{5/2}$, $0g_{9/2}$, and $1d_{5/2}$ orbits for neutrons. 
The degrees of freedom that can be encompassed in this valence space are  very similar 
to the ones  present in the study recalled previously \cite{SDPF-M}, 
and the same kind of phenomena could therefore be described.
The advantage of this valence space is that it contains all 
the physical degrees of freedom important for the description of 
the low-lying properties of these nuclei, it is computationally tractable,
(see precisions hereafter) and it is almost free of center of mass spuriousity 
since its main components, the  $0f_{7/2}\rightarrow0g_{9/2}$ 
excitations, are excluded from the space.

The interaction proposed in this work, denoted hereafter LNPS, is a hybrid one,
based on several sets of realistic two-body matrix elements (TBME). Its main building blocks are:

\begin{itemize}

\item The last evolution of the Kuo-Brown interaction (KB3gr) for the $pf$-shell \cite{KB3GR}; 

\item The renormalized $G$-matrix of Ref. \cite{Jensen95}
with the monopole corrections introduced in \cite{Nowacki-PhD}, for  the remaining
matrix elements involving the $1p_{3/2}$, $1p_{1/2}$, $0f_{5/2}$, and $0g_{9/2}$ neutron orbits. 

\item The $G$-matrix based on the Kahana-Lee-Scott potential \cite{KLS}, for the  matrix elements
involving the $1d_{5/2}$ orbit. 
This potential has been successfully employed in the definition of the recent SDPF-U 
shell model interaction \cite{SDPF-U}
for the description of neutron-rich $sd-pf$ nuclei.

\end{itemize}

In addition, another set of  experimental constraints has been taken into account 
for the final tuning of the monopole Hamiltonian:

\begin{itemize}  

\item The $Z=28$ proton gap  around $^{78}$Ni is inferred from recent experimental
data in  $^{80}$Zn and fixed so as to reproduce the measured $B(E2)$. 
A standard polarization charge of 0.5$e$ has been used in all the calculations
presented in this work. 

\item The size of the $N=50$ neutron gap in $^{78}$Ni has been 
estimated to be $\sim5$ MeV.
The evolution of the neutron gap with the neutron number
is rather independent on the proton number. 
On the contrary, the systematics \cite{GEMO} shows that the 
$0g_{9/2}-1d_{5/2}$ gap in Zr isotopes increases by 3~MeV
when the neutron $0g_{9/2}$ orbital is filled. We assume a similar
behavior for the Ni chain. 
With this assumption, the observed $5^+, 6^+$
states in $^{82}$Ge, which are supposed to be 1p-1h excitations
across the $N=50$ gap, are correctly reproduced \cite{urban-ge82}.

\end{itemize}
Finally, in order to compensate for the absence in our space of the third member of the
quasi-SU3 sequence, the $2s_{1/2}$ neutron orbit, we have increased the 
quadrupole-quadrupole interaction of the neutron $0g_{9/2}$ and  $1d_{5/2}$ orbits by
20\%.

\begin{figure}

\begin{center}

\includegraphics[scale=0.275]{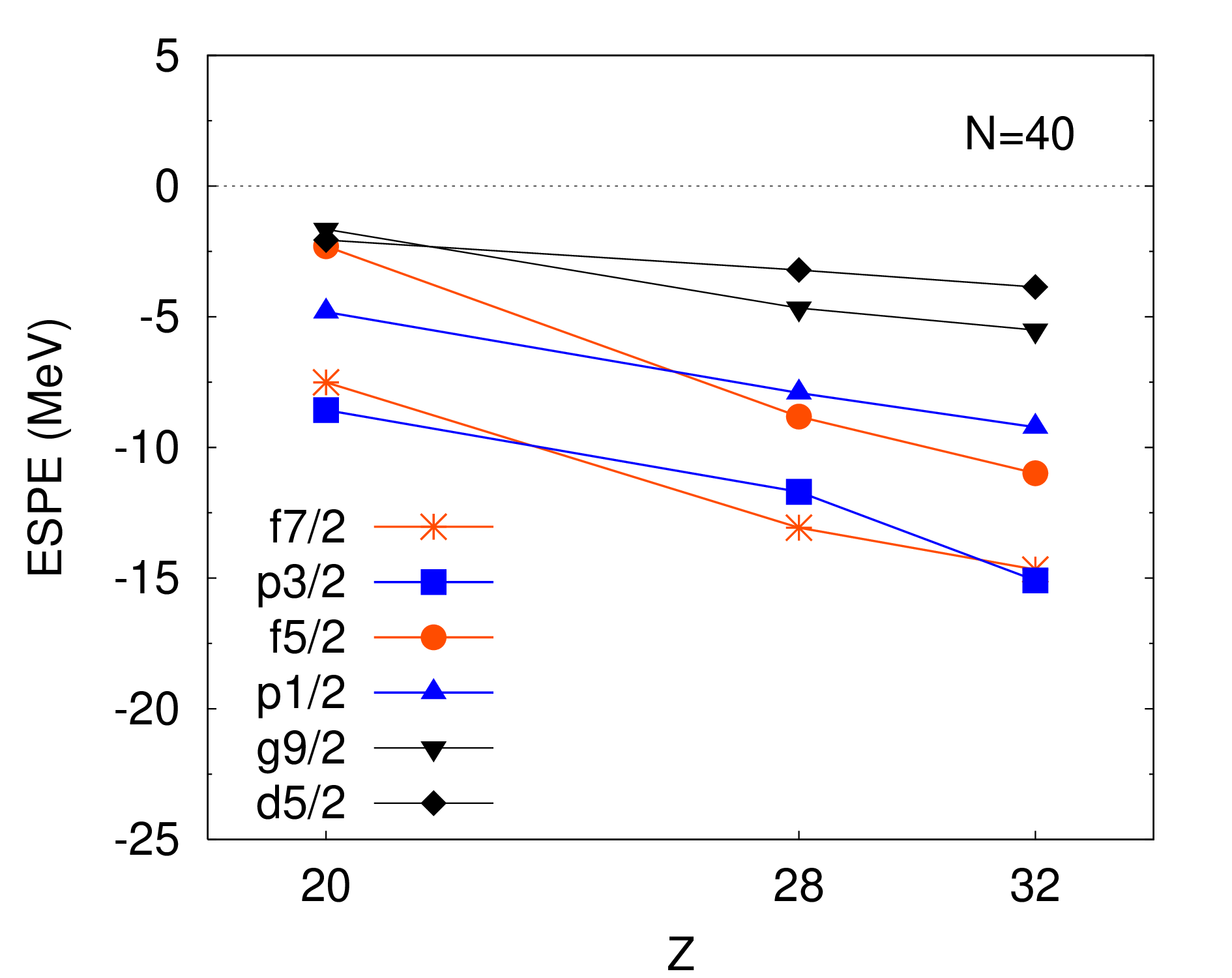}\\
\includegraphics[scale=0.275]{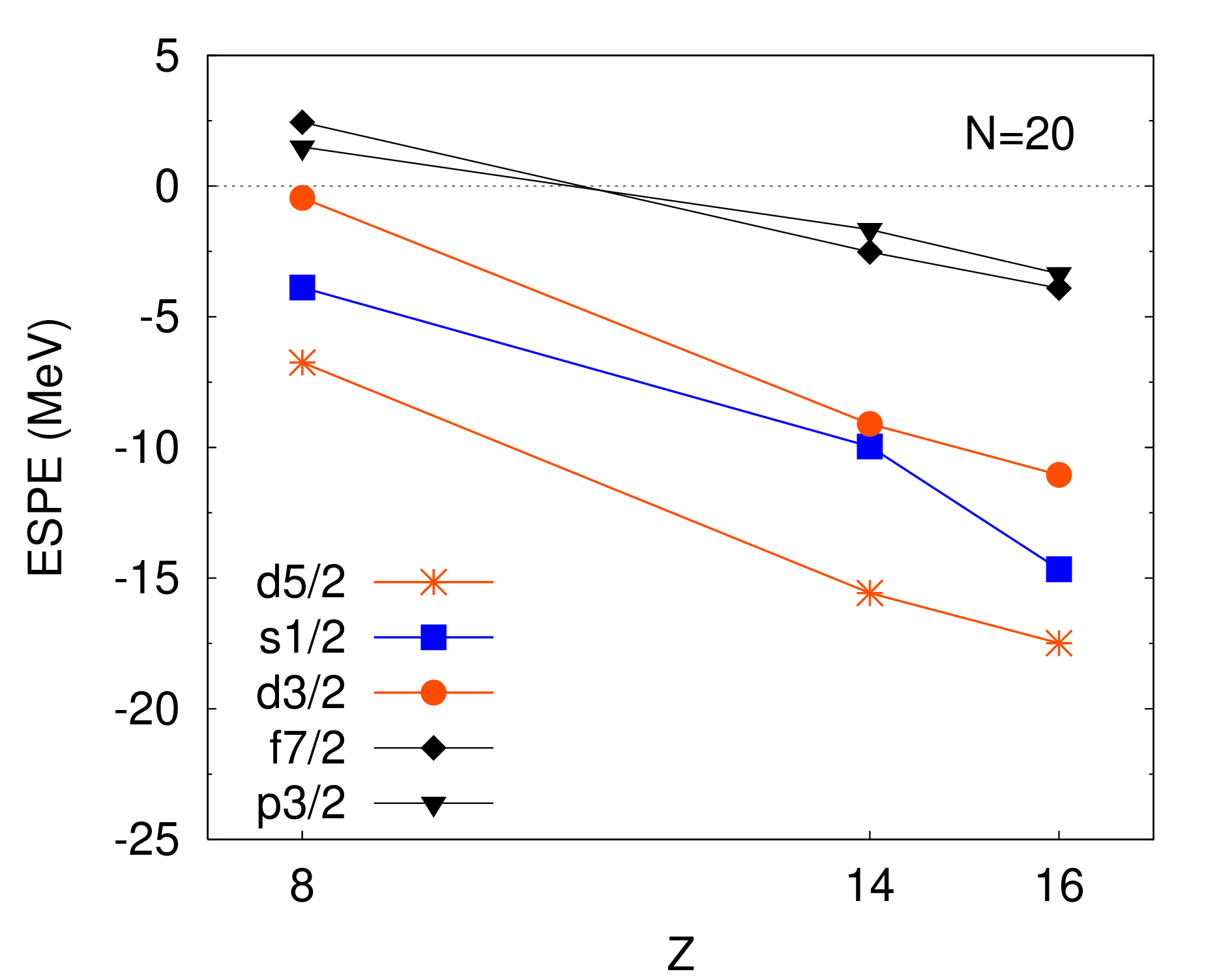}

\caption{(Color online) Neutron effective single particle energies obtained
with the LNPS interaction at $N=40$ (upper part)
and with the SDPF-U interaction at $N=20$ (lower part).\label{fig-espe}}

\end{center}

\end{figure}

The effective single particle energies (ESPE) of the LNPS interaction are shown in Fig. \ref{fig-espe} 
for the neutron orbits at $N=40$, between $^{60}$Ca and $^{72}$Ge.
For comparison, we show in the same figure the neutron ESPE at $N=20$ between 
$^{28}$O and $^{36}$S, calculated with
the SDPF-U interaction \cite{SDPF-U}.
The similarities  are striking. In the $N=20$ case, a reduction of the
neutron $0d_{3/2}-0f_{7/2}$ gap takes place when protons are removed from
the proton $0d_{5/2}$ orbital. This feature, accompanied by the proximity 
of the quadrupole partner neutron orbitals $0f_{7/2}$ and $1p_{3/2}$, is responsible
for the formation of the {\it island of inversion} at $N=20$.
At $N=40$ one observes the same behavior
but for the neutron $0f_{5/2}-0g_{9/2}$ gap when going down from $^{68}$Ni
and the closeness of the quadrupole partners $0g_{9/2}-1d_{5/2}$.

The analogy in the shell evolution between $N=20$ and $N=40$ 
suggests therefore a possibility of 
another {\it island of inversion} below $^{68}$Ni.

As mentioned already, the deformation driving role of the neutron 
$1d_{5/2}$ orbital below $^{68}$Ni was already discussed in the 
shell model calculations of 
Ref. \cite{Caurier2002}. However,
due to the computational limitations 
at that time, the calculations
were performed with a closed neutron $1p_{3/2}$ orbit, therefore using
$^{52}$Ca as a core. This is no longer an issue here:
The present calculations have been carried out including  up to 14 particle-14 hole excitations
across the $Z=28$ and $N=40$ gaps, when it appeared necessary to assure
the convergence of the calculated electromagnetic properties.   
The largest dimensions of the matrices treated here reach
$10^{10}$ in the case of $^{64}$Fe. All
the calculations of this work have been performed using the m-scheme shell model code
{\small ANTOINE} \cite{ANTOINE}.

\section{The properties of the $N=40$ isotones for Z$\leq$28 
and the new region of deformation\label{n40}}

\begin{figure}
\begin{center}
\includegraphics[scale=0.6]{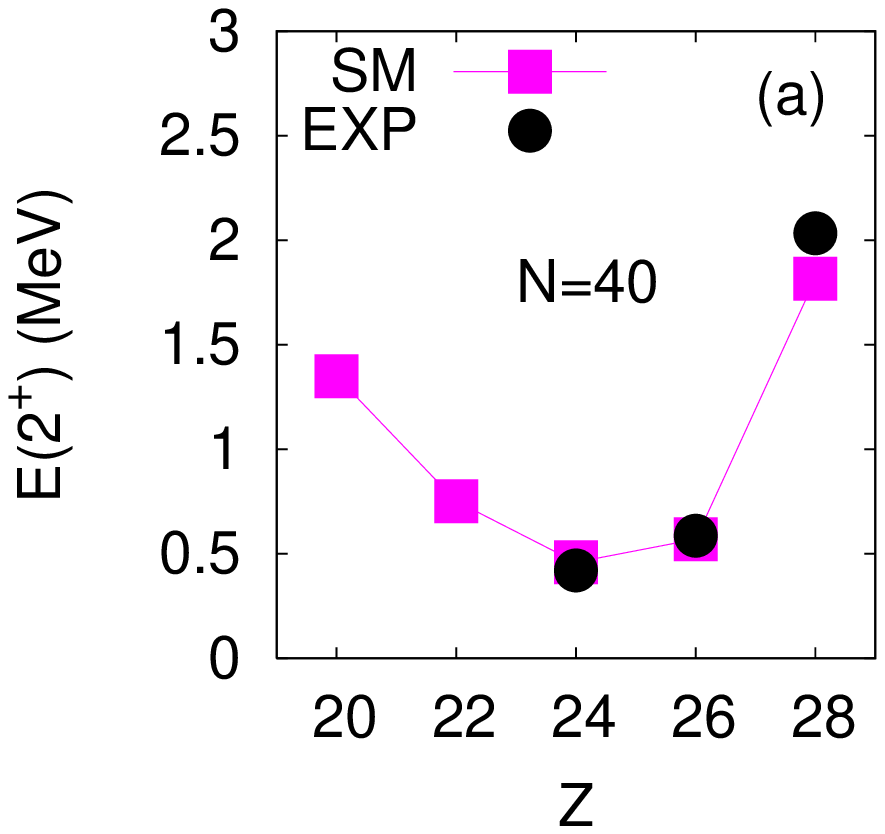}\\
\includegraphics[scale=0.6]{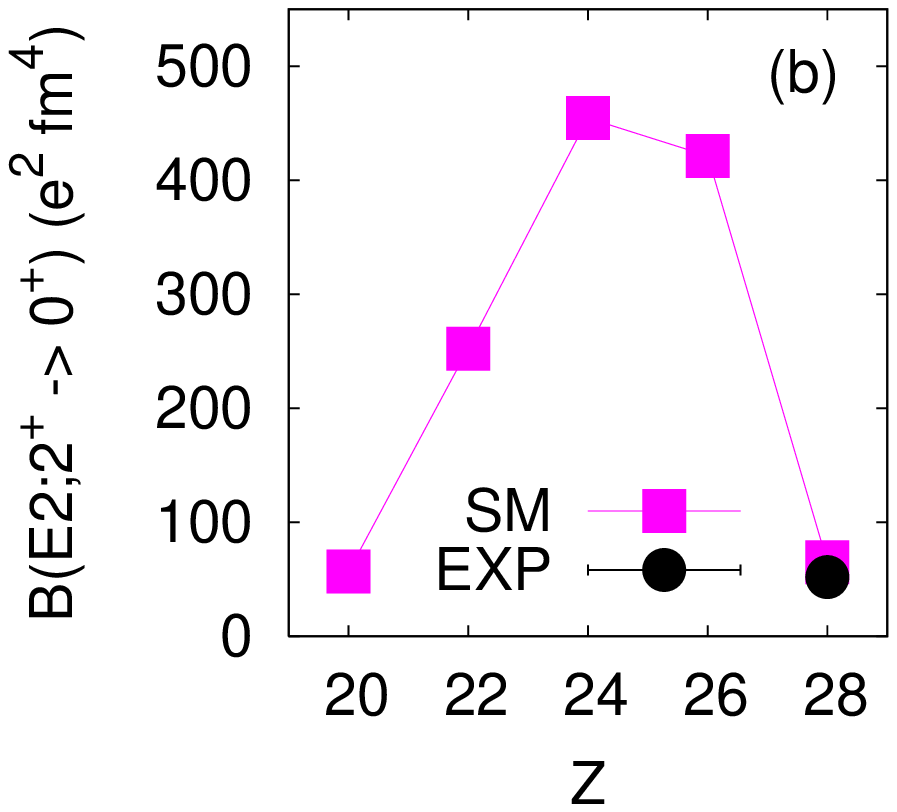}
\caption{(Color online) Evolution of the nuclear structure along the $N=40$ isotonic chain:
(a) theoretical $2^+$ excitation energies and (b) $B(E2; 2^+\rightarrow0^+)$
values, compared to the available experimental data.\label{fig-n40}}
\end{center}
\end{figure}

The change of structure in even-even $N=40$ isotones from $^{68}$Ni down to $^{60}$Ca
is illustrated in Fig. \ref{fig-n40}. In part (a) the evolution
of the excitation energy of the $2^+$ state is plotted while part (b) shows the corresponding evolution
of the $E2$ transition rates.  The available experimental data are satisfactorily reproduced in all cases. 
Our theoretical approach predicts the maximal  deformation 
at the middle of the proton shell, i.e. in $^{64}$Cr, where the calculated $2^+$ energy is
the lowest and the value of the  $B(E2; 2^+\rightarrow 0^+)$ is the largest.

In Table \ref{tab-occ-n40} we list the extra occupancies  
of the two neutron intruder orbitals, $0g_{9/2}$ and $1d_{5/2}$, relative to the normal filling.
The occupation of the $0g_{9/2}$ orbit in $^{68}$Ni is close to~1,
which corresponds to $\approx$50\% of the calculated wave function having
a $J=0$ pair excited across $N=40$ gap. The wave function is however difficult to interpret,
because, in spite of its very large content of excited pairs, the doubly magic component is still substantial. This 
is probably the reason why $^{68}$Ni shows at the same time an increase of the
$2^+$ excitation energy typical for doubly magic nuclei, and no sign of shell
closure in the neutron separation energy.  

The occupation of both neutron intruder
orbits grows rapidly when protons are removed, due to the reduction of the neutron $N=40$
gap shown in Fig. \ref{fig-espe}. The ratio of the $0g_{9/2}$ to $1d_{5/2}$ occupations
also evolves from Ni to Ca, partly as an effect of the level crossing taking place
around $^{62}$Ti, but mostly due to the quasi-SU3 structure of the intruder states.
The very important role of the $1d_{5/2}$ neutron orbit in the build up of collectivity
in this region sheds doubts about the results of Ref. \cite{Kaneko08}, which do not include
this orbit. The more so when one examines the very unrealistic ESPE that they enforce
into their schematic interaction in order to get results close to the experimental data.

In Table \ref{tab-occ-n40}  we list as well the percentages of neutron n-particle n-hole 
excitations in the ground state wave functions. The 4p-4h components are dominant
in Fe, Cr, Ti and Ca, however 2p-2h and 6p-6h contributions are sizable.
The complexity of the wave functions constitutes the main difference 
between $N=40$ and $N=20$ regions. In the latter, nearly pure 2p-2h components
has been shown to dominate the $0^+$ ground states \cite{SDPF-M}.

It should be also pointed out 
that this evolution of the neutron filling and of the particle-hole structure does not mean that the nuclei will
become more and more deformed with decreasing $Z$; the ground-state deformation 
properties result from the total balance between the monopole and the correlation energies 
(mainly of a proton-neutron character).
In Table \ref{tab-occ-n40}, we  list also these correlation energies extracted from
the multipole Hamiltonian along the $N = 40$ line.
Indeed,  the correlation energy, reflecting the deformation,
increases from Ni to Cr, where it reaches its maximum, and then diminishes toward Ca. 
The transition between Ni and Cr is not gradual: The removal of two protons already provokes
and abrupt change, from 
spherical to a  strongly deformed  prolate shape.

\begin{table}
\caption{Occupation of the neutron intruder orbitals 
and percentage of particle-hole excitations across the $N=40$ gap
in the ground states
of the $N=40$ isotones. The last column contains the correlation energies
evaluated for these states.\label{tab-occ-n40}}
\begin{tabular}{cccccccc}
\hline
\hline
Nucleus & $\nu g_{9/2}$ & $\nu d_{5/2}$ &0p0h & 2p2h & 4p4h & 6p6h & E$_{corr}$\\
\hline
$^{68}$Ni &0.98 &0.10  &55.5 &35.5 &8.5 &0.5  &-9.03\\
$^{66}$Fe &3.17 &0.46  &1 &19 &72 &8  &-23.96\\
$^{64}$Cr &3.41 &0.76  &0 &9  &73 &18 &-24.83\\
$^{62}$Ti &3.17 &1.09  &1 &14 &63 &22 &-19.62\\
$^{60}$Ca &2.55 &1.52  &1 &18 &59 &22 &-12.09\\
\hline
\hline
\end{tabular} 
\end{table}

\section{The evolution of the deformation along the isotopic chains\label{ich}}

\begin{figure}  
\begin{center}
\includegraphics[scale=0.6]{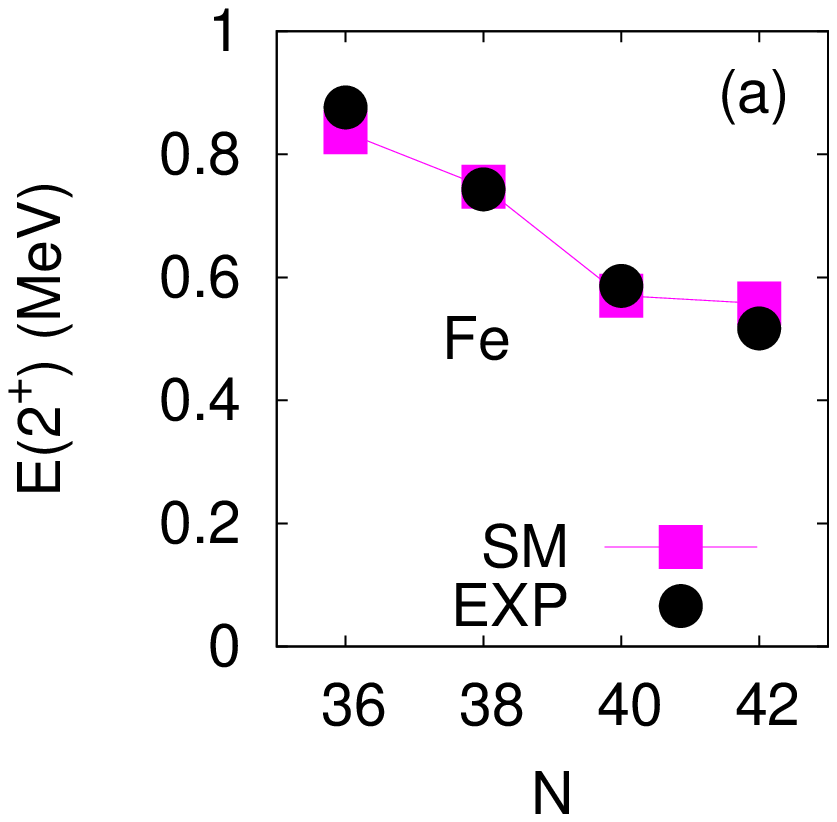}\\
\includegraphics[scale=0.6]{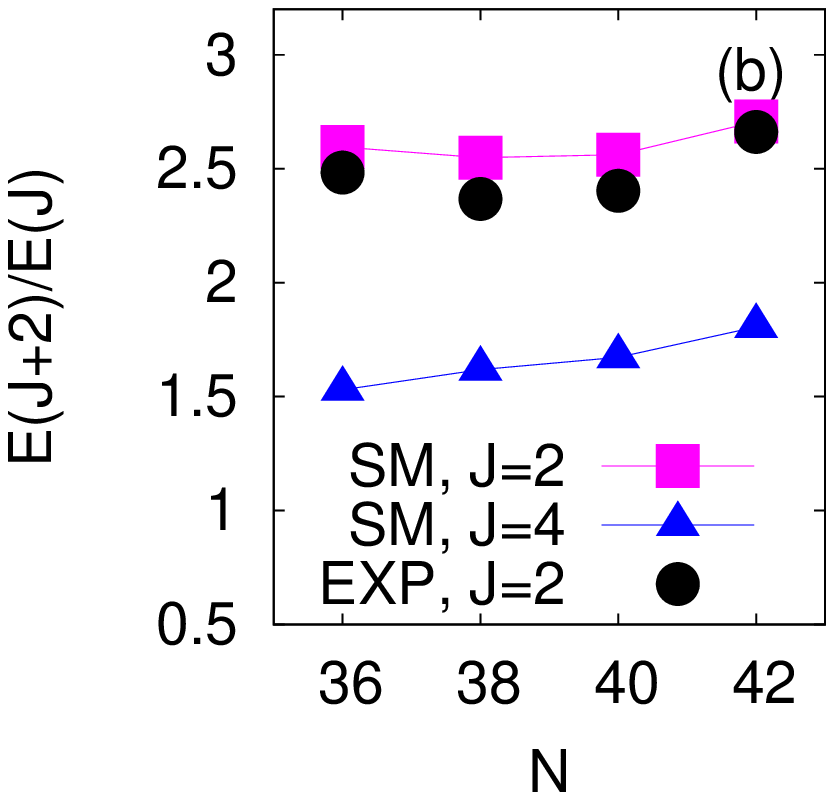}\\
\includegraphics[scale=0.6]{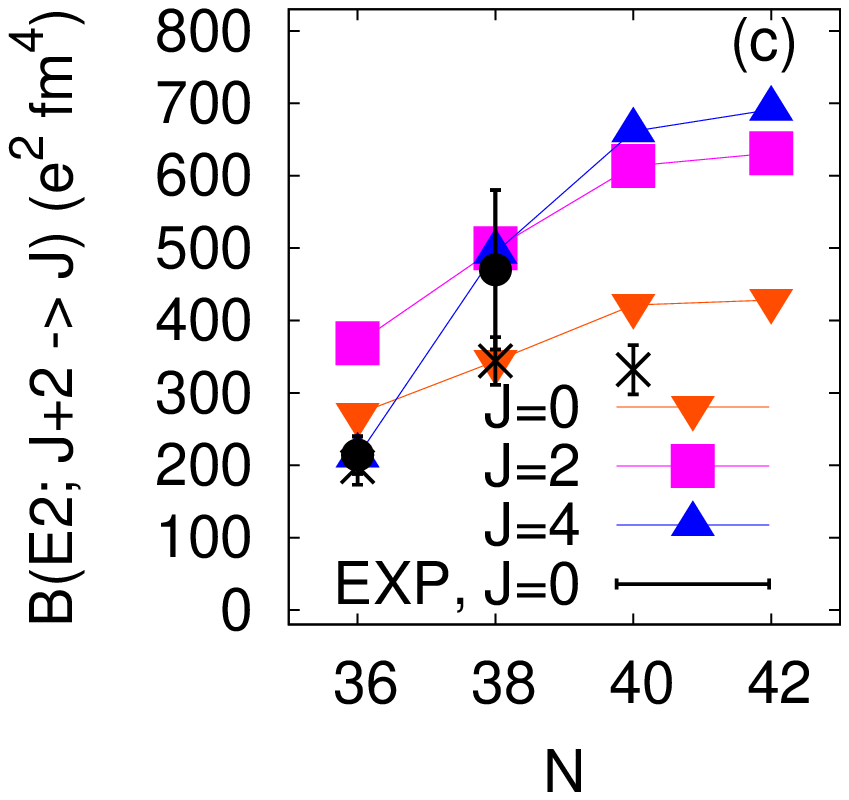}\\
\includegraphics[scale=0.6]{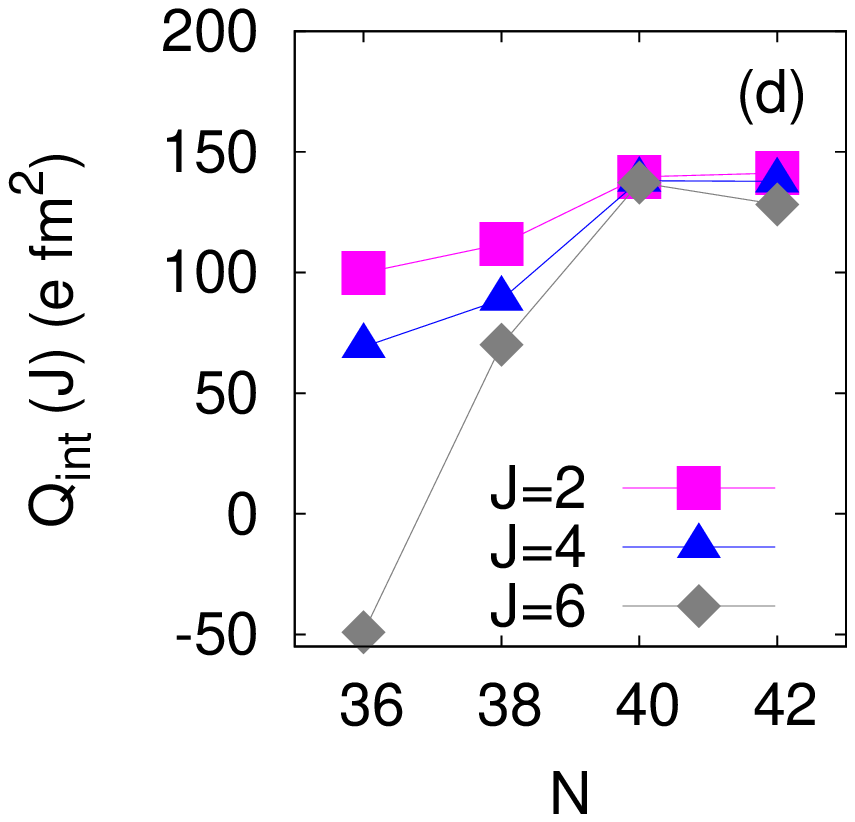}
\caption{(Color online) Theoretical results along the iron isotopic chain 
in comparison with the available experimental data: the excitation energies
of the $2^+$ states are shown in panel (a), in panel (b) we present
the ratio of energies of $E(J+2)/E(J)$, the $B(E2)$ transition rates
are plotted in panel (c) and the calculated intrinsic quadrupole moments in panel (d).
Two experimental sets of the $B(E2)$ values are shown: from Ref. \cite{Ljungvall} (black dots)
and from Ref.~\cite{Rother} (crosses). \label{fig-fe}}
\end{center}
\end{figure}

The isotope chains provide a very illustrative picture of the evolution 
of deformation in this region. 
In Fig. \ref{fig-fe} we show the results obtained for the iron chain.
In the panel (a) the excitation energies of the $2^+$
states are compared to the available experimental data.
In panel (b) we show the ratio of the excitation energies $E(4^+)/E(2^+)$ and $E(6^+)/E(4^+)$, 
which make it possible to recognize whether a nuclear spectrum  
is close to that of the rigid rotor or not. Let us remind that the perfect rotational limit
would require $E(4^+)/E(2^+)$=3.33 and $E(6^+)/E(4^+)$=2.1.
Panel (c) depicts the theoretical and experimental $B(E2)$ transition probabilities along the yrast bands. 
The measured 
$B(E2; 2^+\rightarrow0^+)$ values from Ref.~\cite{Ljungvall} are plotted with dots and
from Ref.~\cite{Rother} with crosses. Finally, 
the intrinsic quadrupole moments
derived from the calculated spectroscopic ones are shown in part (d).
To establish a connection between the laboratory  and the intrinsic frames
we use the relations:

\begin{equation}
Q_{int}=\frac{(J+1)(2J+3)}{3K^2-J(J+1)}Q_{spec}(J),\,\,\,\, K\ne1, \label{eq1}
\end{equation}

and

\begin{equation}
B(E2,J\rightarrow J-2)=\\ \frac{5}{16}e^2|\langle JK20|J-2,K\rangle|^2Q_{int}^2 \label{eq2}
\end{equation}

\noindent  for $K\ne\frac{1}{2}, \;1$.

\begin{figure}  
\begin{center}
\includegraphics[scale=0.6]{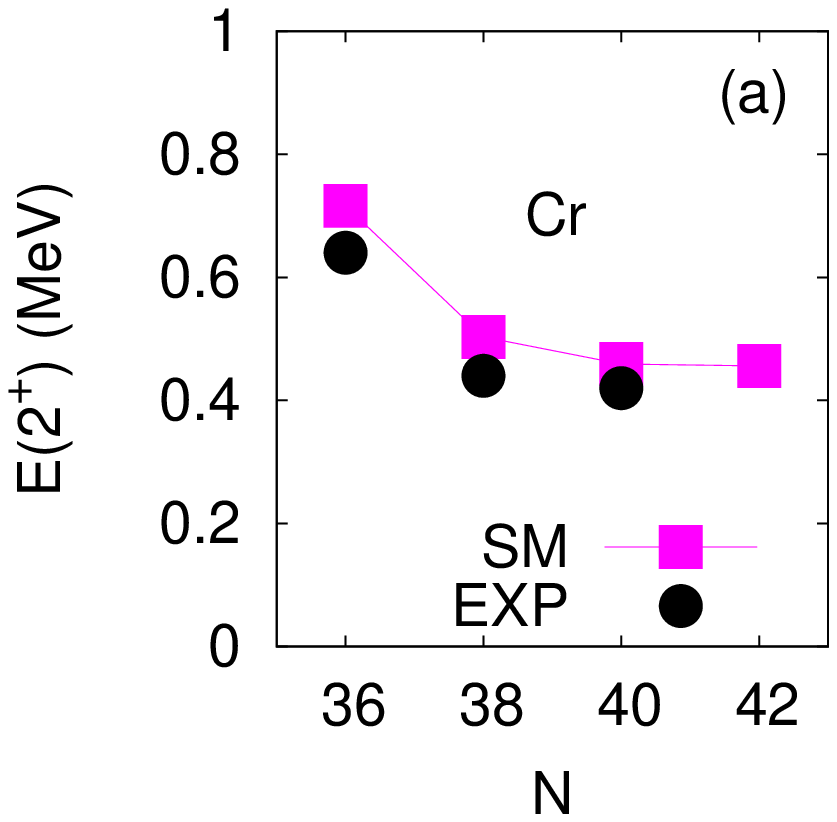}\\
\includegraphics[scale=0.6]{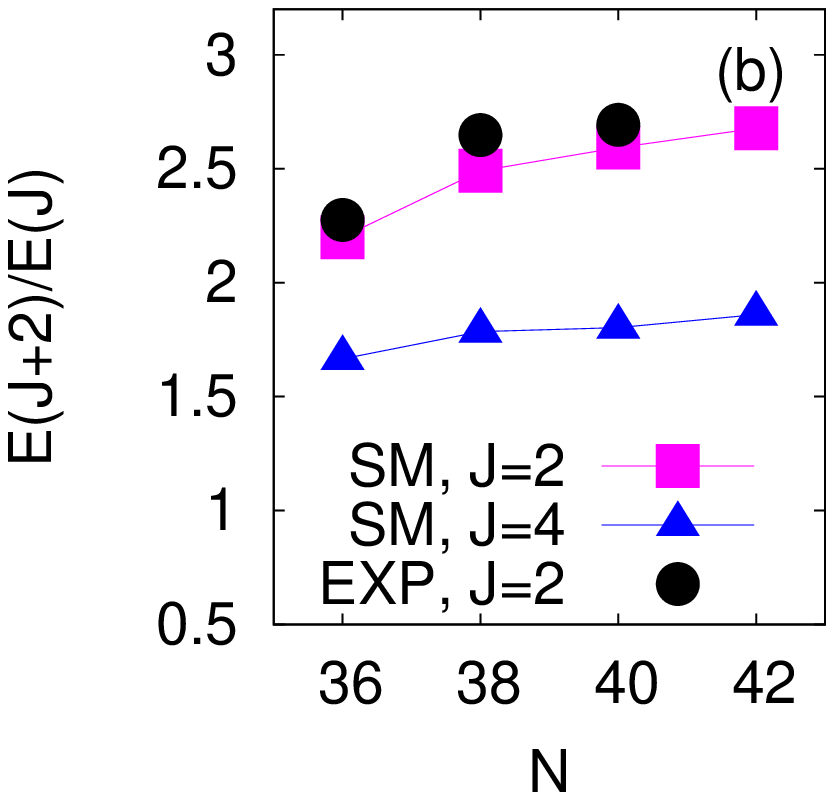}\\
\includegraphics[scale=0.6]{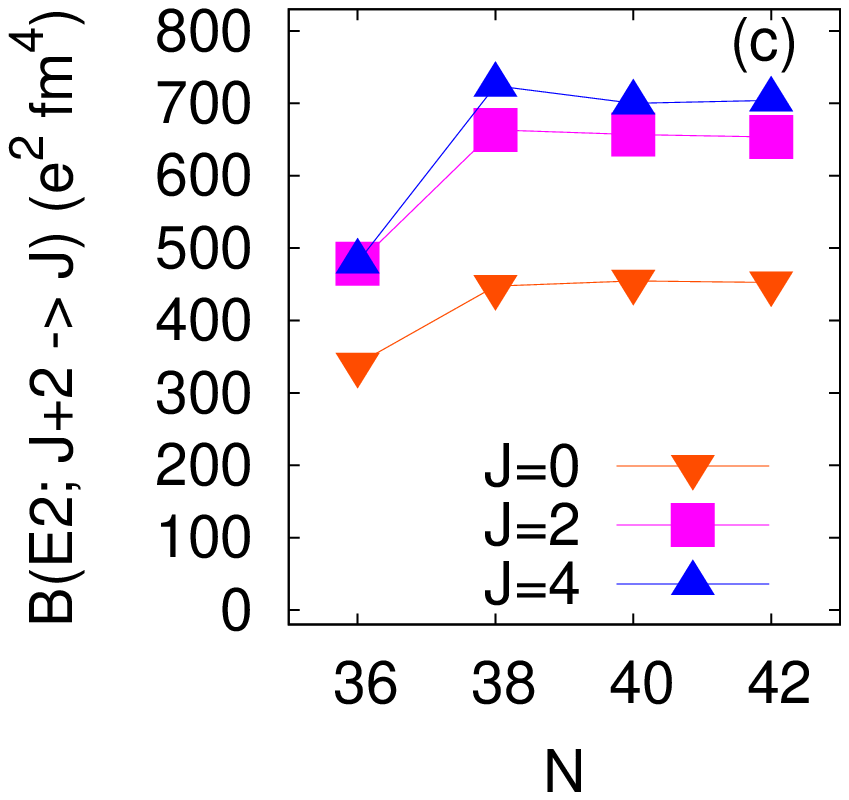}\\
\includegraphics[scale=0.6]{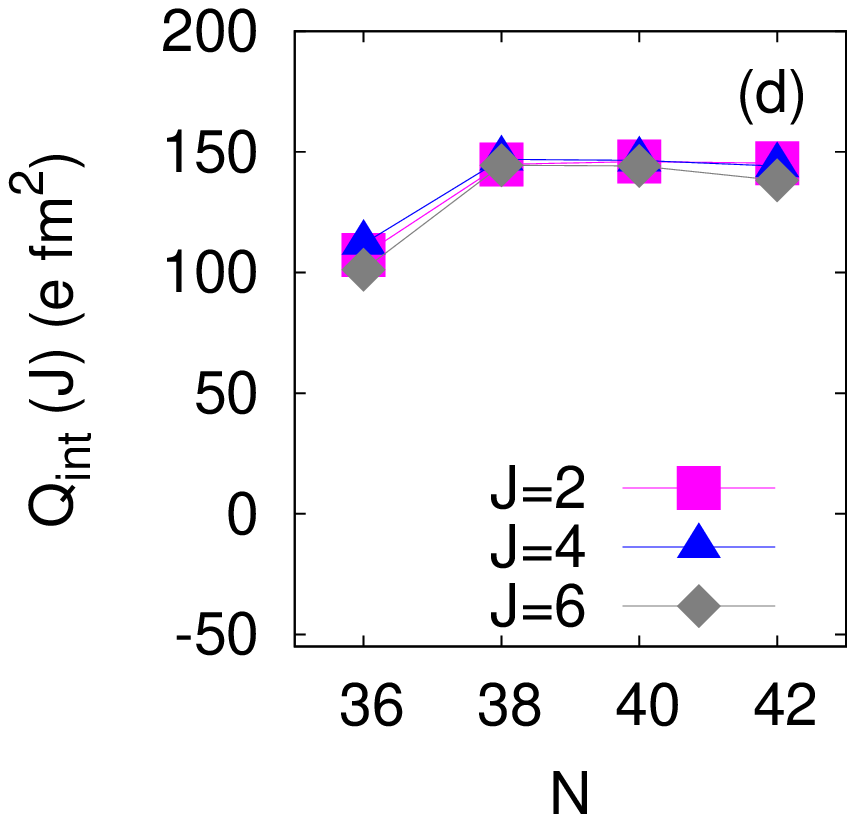}
\caption{(Color online) The same as in Fig. \ref{fig-fe} but for the chromium isotopes.\label{fig-cr}}
\end{center}
\end{figure}

The same properties  are plotted in Fig. \ref{fig-cr}  for the chromium chain. 
An excellent agreement with the experiment is found for all excitation energies.

The known $B(E2)$ transition rates are well  reproduced within the error bars as well.
Comparing the results for both isotopic chains it can be seen  that the onset of deformation occurs
at different neutron number. In the chromium chain the intrinsic quadrupole moment stays constant
along the yrast the band already at $N=38$. This is one  of the fingerprints of 
a good rotor behavior. The iron isotopes undergo the transition
at $N=40$. We have also verified that the intrinsic quadrupole moments
obtained from the spectroscopic ones (Eq. (\ref{eq1})) are nearly equal to those obtained
from the transition probabilities according to Eq. (\ref{eq2}) in $^{62,64,66}$Cr and
$^{66,68}$Fe. We obtain a value of $Q_{int}\sim150e$fm$^2$ in $^{62,64,66}$Cr
which corresponds to $\beta\sim0.35$. The deformation of $^{66,68}$Fe is slightly lower, 
with $Q_{int}=145e$fm$^2$ ($\beta\sim0.3$). 
In the lighter Fe and Cr nuclei, the discrepancies between the calculated quadrupole moments and those 
obtained in the rotational scheme from the $B(E2)$'s are much larger.

Let us further discuss the structure of the calculated states.
In Table \ref{tab-occ} we show the occupancies of the neutron intruder orbitals
in the ground states of the chromium and iron chains. In the case of 
the $0g_{9/2}$ orbit 
the extra occupancy is reported, i.e. the difference between the value obtained in
the configuration mixing calculation and the one that corresponds to the normal  filling.
The neutron intruder occupations increase with $N$ in both isotopic chains, however the
absolute occupancies are larger in the chromium chain. As mentioned already for the $N=40$ nuclei,
the strong deformation in the Cr chain is not only due to the increased population
of the $0g_{9/2}-1d_{5/2}$ doublet, but  to the strong proton-neutron correlations which tend to be
maximal when four protons are active in the $pf$ shell as well.

\begin{table}
\begin{center}
\caption{Extra occupations of neutron $0g_{9/2}$ and $1d_{5/2}$ orbitals
in the ground states of the chromium and iron chains.\label{tab-occ}}
\begin{tabular}{cccc}
\hline
\hline
Nucleus & $N$ & $\nu 0g_{9/2}$ & $\nu 1d_{5/2}$\\ 
\hline
$^{62}$Fe &36 &0.95& 0.12\\
$^{64}$Fe &38 &2.0 & 0.27\\
$^{66}$Fe &40 &3.22& 0.51\\
$^{68}$Fe &42 &2.30& 0.62\\
\hline 
$^{60}$Cr &36 &1.55 & 0.31\\
$^{62}$Cr &38 &2.77 & 0.66\\
$^{64}$Cr &40 &3.41 & 0.76\\
$^{66}$Cr &42 &2.28 & 0.90\\
\hline
\hline
\end{tabular}
\end{center}
\end{table}

\begin{figure}  
\begin{center}
\includegraphics[scale=0.6]{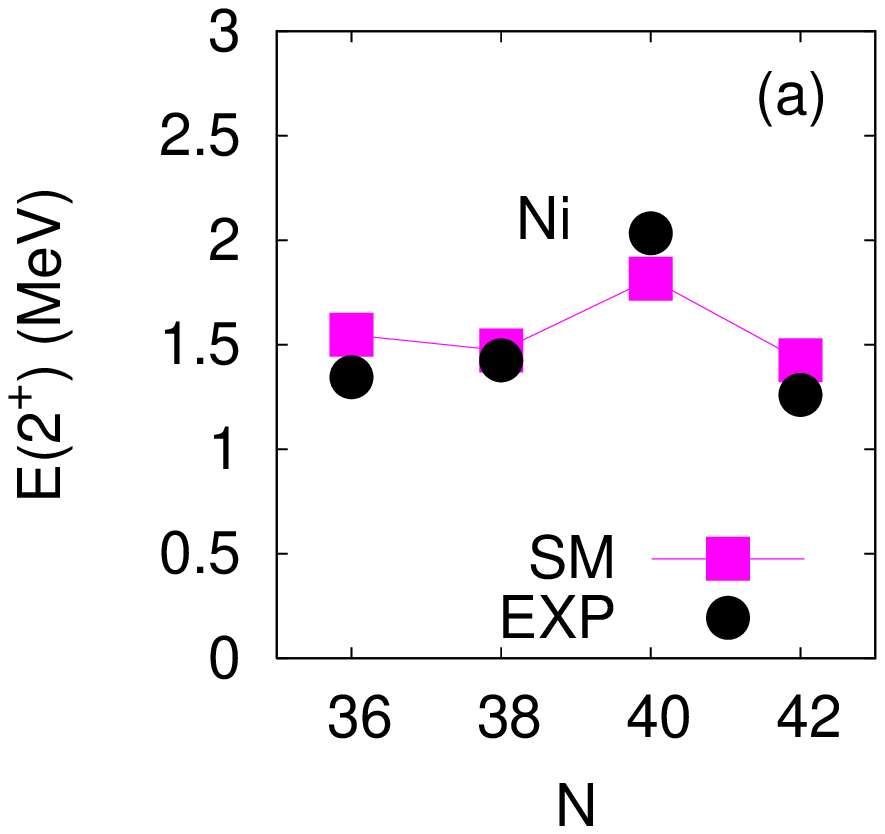}\\
\includegraphics[scale=0.6]{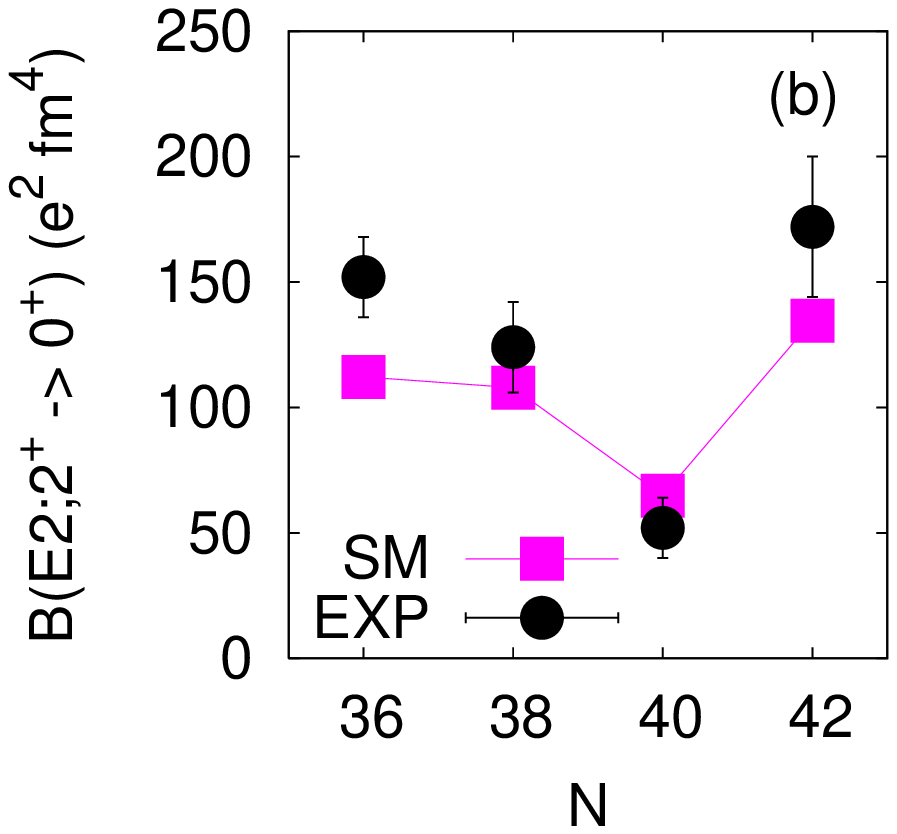}
\caption{(Color online) Theoretical versus experimental energies and transition rates
of the nickel isotopes in the vicinity of $N=40$. \label{fig-ni}}
\end{center}
\end{figure}

In Fig. \ref{fig-ni} we show also the evolution of the collectivity in the nickel
chain: in (a) the theoretical energies of the first excited
$2^+$ states are compared with experimental ones, while in (b) we show the 
calculated transition rates in comparison to the available data.
The agreement in the calculated energies is very good for all nickels.
Concerning transition rates, the model reproduces well the systematics
with the minimum in $^{68}$Ni and the rapid increase of collectivity in $^{70}$Ni.
However, the calculated value is closer to the lower tip of the error bar. 
The transition rate in $^{64}$Ni is underestimated.
In this case we know that a better agreement
with experiment can be obtained in a full $pf$ shell calculation
as the neutron excitations from $0f_{7/2}$ orbital are here more important than
those through the $N=50$ gap and, as expected,  the occupation of the $1d_{5/2}$
orbit remains close to zero in the calculations. 

Finally, let us note the recent work on $^{68}$Ni \cite{Pauwels:2010xa}, where
the excited $0^+$ states have been investigated and a candidate for a proton 2p-2h
intruder has been proposed at an excitation  energy of 2.2~MeV.
We have calculated the excited $0^+$ states in our model space and we
obtain the first excited $0_2^+$ state at an energy of 1.2~MeV
and the second one $0_3^+$ at an energy of 2.4~MeV, in good agreement with experiment.
The too low excitation energy of the first state can be a result of the delicate 
mixing between the $0_1^+$ ground state and the excited $0_2^+$ state.
These two states are characterized by similar
proton occupancies with leading 0p-0h (neutron) configuration for the $0_1^+$ ground state
and 2p-2h (neutron) configurations for the $0_2^+$.
For  the $0_3^+$ state, located at 2.4~MeV, the striking feature is that the dominant
proton configuration has exactly two $0f_{7/2}$ protons less than the ground state, $i. e.$
a pure 2p-2h proton configuration.

The total quadrupole sum rules for these states amount to  904, 1162 and 2025 $e^2$fm$^4$ 
for the ground state, the $0_2^+$, and the  $0_3^+$ states respectively. This shows in particular that 
the $0_3^+$ state carries moderate deformation, despite of its proton intruder nature.

\section{Conclusions\label{con}}
The aim of this work was to study the rapid onset of collectivity below $^{68}$Ni, suggested
by the experimental evidence, which is a great challenge for any theoretical model. Here
we have presented shell model calculations in a large valence space
including the $pf$ shell for protons and the $0f_{5/2}$, $1p_{3/2}$, $1p_{1/2}$, $0g_{9/2}$, and $1d_{5/2}$ orbitals for neutrons.
The effective interaction for this model space has been build up  using sets of realistic TBME
as a starting point and 
applying monopole corrections thereafter.
The progresses in algorithms and computer power have made it possible to achieve
the largest shell model diagonalizations 
in this region of nuclei up to date.  

The present calculations have been performed in Ni, Cr, Fe, Ti and Ca
isotopes around  $N=40$. We have found a satisfactory agreement between the theoretical results
and the available data, for both excitation energies and transition rates. In particular, it has been possible
to describe correctly the rapid onset of collectivity below $^{68}$Ni 
without any collapse of the spherical gaps.
 
It has been also shown that the onset of deformation develops at $N=40$ in the iron chain 
and already
at $N=38$ in chromium isotopes. The maximum deformation is predicted for the chromiums
which exhibit features typical of rotational nuclei. 
The continuous advances in the experimental
side, with particular regard to the
transition probability measurements, will be a stringent test for these theoretical predictions.

The calculated wave functions of the deformed ground states were shown to contain large
amounts of many particle-many hole configurations, with around four neutrons
occupying intruder orbitals $0g_{9/2},1d_{5/2}$. 
We have also related the observed shape change with the underlying evolution of the spherical mean-field,
which bears many similarities  with the  one at  the $N=20$ {\it island of inversion}.

{\bf Acknowledgments.} This work is partly supported  by a grant of
                       the Spanish Ministry of Science and Innovation MICINN
                       (FPA2009-13377), by the IN2P3(France)-CICyT(Spain)
                       collaboration agreements, by the Spanish
                       Consolider-Ingenio 2010 Program CPAN
                       (CSD2007-00042) and by the Comunidad de Madrid
                       (Spain), project HEPHACOS S2009/ESP-1473.



\end{document}